\font\bigbf=cmbx12 scaled\magstep2
\font\bigifont=cmti10
\font\bigrfont=cmr12
\font\srfont=cmr8

\vsize =23.5 truecm
\hsize = 15.0 truecm
\hoffset = .2truein
\baselineskip 1.0\baselineskip

\centerline{\bigbf Phenomenological Lambda-Nuclear Interactions}
\vskip 1.0 truecm

\centerline{\bigrfont Rita Sinha}

{\baselineskip= 16 truept
\centerline{\bigifont Department of Physics, Universiti Putra Malaysia, UPM 43400
Serdang, Selangor D.E., Malaysia.
}

\vskip 0.8 truecm

\centerline{\bigrfont {\footnote{$^\dagger$}
{\srfont On leave from Department of Physics, Jamia Millia Islamia, New Delhi, 110025 India.}}
Q. N. Usmani}
{\baselineskip= 16 truept
\centerline{\bigifont Department of Physics, Universiti Putra Malaysia, UPM 43400
Serdang, Selangor D.E., Malaysia.
}
{\baselineskip= 16 truept
\centerline{\bigifont and Theoretical Studies Laboratory, Institute of Advanced Technology, Universiti Putra Malaysia, UPM 43400 }
\centerline{\bigifont Serdang, Selangor D.E., Malaysia.}
}

\vskip 0.8 truecm
\centerline {\bigrfont B. M. Taib}

{\baselineskip= 16 truept
\centerline{\bigifont Institute of Multimedia and Software, Universiti Putra Malaysia, UPM 43400  Serdang, Selangor D.E., Malaysia.
}

\vskip 0.8 truecm

\centerline{ \bf  [ABSTRACT]}

Variational Monte Carlo calculations for ${_{\Lambda}^4}H$ (ground and excited states) and ${_{\Lambda}^5}He$ are performed to decipher information on ${\Lambda}$-nuclear interactions. Appropriate operatorial nuclear and ${\Lambda}$-nuclear correlations have been incorporated to minimize the expectation values of the energies. We use the Argonne $\upsilon_{18}$ two-body {\it NN} along with the Urbana $IX$ three-body {\it NNN} interactions. The study demonstrates that a large part of the splitting energy in ${_{\Lambda}^4}H$ ($0^+-1^+$) is due to the three-body ${\Lambda}${\it NN} forces. $_{\Lambda}^{17}O$ hypernucleus is analyzed using the  {\it s}-shell results. $\Lambda$ binding to nuclear matter is calculated within the variational framework using the Fermi-Hypernetted-Chain technique. There is a need to correctly incorporate the three-body ${\Lambda}${\it NN} correlations for  $\Lambda$ binding to nuclear matter.

\vskip 12 truept
\noindent PACS number(s): 21.80.+a, 21.10.Dr, 13.75.Ev, 24.10.Lx, 21.65.+f

\vskip 1.0 truecm

\centerline{ \bf \S 1. Introduction}
\vskip 12 truept

The study of the response of a many-body system to a hyperon gives insight into the structure
of baryon-baryon interactions. The binding energy data of light {\it s}-shell hypernuclei
provide a unique opportunity to know more about the $\Lambda$-nuclear interactions,
particularly on their spin-dependence. In the past, basically two approaches have been
followed. The first one involves Brueckner-Hartree calculations using Nijmegen {\it YN}
potential with and without higher order correction to single-particle energies[1,2].
This method uses the large $\Lambda$$N$$\rightarrow$$\Sigma$$N$ coupling which gives
considerably lower binding energy for ${_{\Lambda}^5}He$. Any attempt to correct this leads to
poor agreement with the scattering data. The second method is primarily based on reliable
variational techniques, mostly using simplified {\it NN} interactions[3,4]. We follow this
approach but use realistic Argonne $\upsilon_{18}$ {\it NN} interaction[5] along with
Urbana IX three-body {\it NNN} interaction[6,7]. The phenomenological approach
we follow is consistent with meson-theoretic models as well as available low energy scattering
data. The $\Lambda$$N$$\rightarrow$$\Sigma$$N$ coupling is effectively taken care of by
inclusion of the phenomenological $\Lambda${\it NN} potential. The $\Lambda${\it NN} potential
consists of both the dispersive and two-pion-exchange (TPE) kind as employed in previous
studies[4]. The hypernuclei considered in this work are ${_{\Lambda}^4}H$,
${_{\Lambda}^4}H^*$ ( * on ${_{\Lambda}^4}H$ refers to the excited 1$^+$ state), and
${_{\Lambda}^5}He$.

The interaction parameters which we find based on our {\it s}-shell results are later used to
make estimates of the binding energy of $_{\Lambda}^{17}O$. We also study the
$\Lambda$ binding to nuclear matter by using the {\it Fermi-Hypernetted-Chain (FHNC)}
technique[4, 8]. This study gives an indication of the implications of our {\it s}-shell
results on heavier hypernuclear systems.

In section 2 we describe the Hamiltonian used in this work. Section 3 gives the wave function
and approach. In section 4 we discuss the results and finally in section 5 we give the
conclusion and comments.

\vskip 1.0 truecm
\centerline{\bf \S 2. The Hamiltonian}
\vskip 12 truept

The complete hypernuclear Hamiltonian consists of the nuclear Hamiltonian $H_N^{A-1}$ and
the lambda Hamiltonian $H_{\Lambda}$. The nuclear Hamiltonian $H_N^{A-1}$ is given by
$$
H_N^{A-1} = -\sum_{i=1}^{A-1} { {\hbar^2} \over {2m_i}} \nabla_i^2 + \sum_{i<j}^{A-1}V_{ij}
+ \sum_{i<j<k}^{A-1}V_{ijk} \,\,, \eqno(2.1)
$$
where $V_{ij}$ and $V_{ijk}$ are the two-nucleon {\it NN} and three-nucleon {\it NNN}
potentials, respectively, and $m_i$ is the mass of the nucleon.

The two-body {\it NN} interaction employed here is the {\it Argonne }$\upsilon_{18}$
interaction[5]. The first fourteen operator components of this model are charge-
independent and are an updated version of the {\it Argonne }$\upsilon_{14}$ potential[9].
Three additional charge-dependent and one charge-asymmetric operators are added along with
a complete electromagnetic interaction, containing the Coulomb, Darwin-Foldy, vacuum
polarization, and magnetic moment terms with finite-size effects. The potential has been fit
directly to the Nijmegen {\it{pp}} and {\it{np}} scattering data base[10,11], low-energy {\it nn} scattering parameters, and deuteron binding energy.
For the three-body {\it NNN} potential we use the Urbana model[6,7] consisting of
the two-pion-exchange (TPE) part of Fujita and Miyazawa[12] and a repulsive
phenomenological spin-isospin independent term. We have used the Urbana $IX$
model[7] of the interaction where the values of the strength parameters are used in
conjunction with the {\it Argonne } $\upsilon_{18}$ interaction.

The lambda Hamiltonian $H_{ \Lambda}$ is given by
$$
H_{\Lambda} = -{ {\hbar^2} \over{2m_{\Lambda}}} \nabla_{\Lambda}^2 + \sum_{i=1}^{A-1}
V_{i{\Lambda}} + \sum_{i<j}^{A-1}V_{ij{\Lambda}} \,\,, \eqno(2.2)
$$
where $V_{i{\Lambda}}$ and $V_{ij{\Lambda}}$ are the two-body ${\Lambda}${\it N} and
three-body ${\Lambda}${\it NN} potentials, respectively, and $m_{\Lambda}$ is the mass of the
${\Lambda}$ particle. The first terms of Eqs.(2.1) and (2.2) pertain to the total kinetic
energy of the nucleons and ${\Lambda}$, respectively.

The two-body $\Lambda${\it N} potential $V_{\Lambda N}$ includes a central potential[4]
of the same form for the singlet and triplet spin states. These have a theoretically
reasonable attractive tail due to the TPE in accord with Urbana type potentials
[13] with spin- and space-exchange terms.

$$
V_{\Lambda N} = \bigl [ (V_c(r) - \bar V)\, (1- \epsilon + \epsilon P_x) + {1 \over 4}
V_{\sigma} \,\vec \sigma_{\Lambda} \cdot \vec \sigma_N  \bigr ] T_{\pi}^2(r) \,\,, \eqno(2.3)
$$
where $\bar V$ and $V_{\sigma}$ are the spin-average and spin-dependent strengths,
respectively. $P_x$ is the Majorana space-exchange operator, $\epsilon$ is the corresponding
exchange parameter, $V_c$ is the Woods-Saxon repulsive core[4], and $T_{\pi}$ is the
one-pion-exchange (OPE) tensor potential shape modified with a cutoff. Further details can be found in Ref. [4].

In this study we consider potential parameters that are consistent with low energy
$\Lambda{\it p}$ scattering data that essentially determine the value of spin-average
strength $\bar V=6.15 \pm 0.05$ MeV [4].

For hypernuclei with zero-spin core nuclei, such as $_{\Lambda}^5He$, the major contribution
arises from the spin-average strength ${\bar V}$ while the spin component contributes very
little. The spin dependence $V_{\sigma}$ is assumed to be positive, which is consistent with
hypernuclear spins of mass 4 systems. We find that for the {\it{s}}-shell hypernuclei
$(A \leq 5)$ the {\it{s}}-state interaction is dominant but the higher partial wave
interactions, in particular, the {\it{p}}-state, also make a small but significant
contribution contrary to earlier studies[14]. The importance of the {\it{p}}-state
contribution becomes significant due to the $\Lambda$-nuclear correlations.

Studies on hypernuclei have shown that it is necessary to include a three-body
${\Lambda}${\it NN} interaction in the Hamiltonian. We consider phenomenological
${\Lambda}${\it NN} forces of the dispersive (spin-dependent and spin-independent) as well as
the TPE kind$^{15)}$, which arise from the suppression of $\Sigma$, $\Delta$, $\ldots$
degrees of freedom by the medium, that is, the second nucleon.

The dispersive kind has a spin dependence that is given by
$$
V_{\Lambda NN}^{DS}(r_{ij {\Lambda }}) = W_o T_{\pi}^2(r_{i \Lambda}) \, T_{\pi}^2(r_{j
\Lambda}) \lbrack {1 + {1 \over 6}
\vec \sigma_{\Lambda} \cdot (\vec \sigma_{i} + \vec \sigma_{j})} \rbrack \,\,. \eqno(2.4)
$$

The TPE part of the interaction is given by[15]
$$
W_p = -{1 \over 6} \, C_p (\vec \tau_i \cdot \vec \tau_j) \{ X_{i \Lambda}, X_{j \Lambda} \}
Y(r_{i \Lambda}) \, Y(r_{j \Lambda}) \,\,, \eqno(2.5)
$$
where X$_{k\Lambda}$ is the OPE operator given by
$$
X_{k\Lambda} = ( \vec \sigma_{k} \cdot \vec \sigma_{\Lambda}) + S_{k \Lambda}(r_{k \Lambda})
\,
T_{\pi}(r_{k \Lambda}) \eqno(2.6)
$$
{\rm with}
$$
S_{k \Lambda}(r_{k \Lambda}) = {3 \, {(\vec \sigma_{k} \cdot r_{k \Lambda})
(\vec \sigma_{\Lambda} \cdot r_{k \Lambda})} \over {r_{k \Lambda}^2}} - (\vec \sigma_{k}
\cdot
\vec \sigma_{\Lambda}) \,\,. \eqno(2.7)
$$
\noindent In Eq.(2.5) {$\{ \}$} represents the anticommutator term.  Y$_{\pi}(r_{k \Lambda})$
and T$_{ \pi}(r_{k \Lambda})$ are the usual Yukawa and tensor functions, respectively, with
pion mass $\mu=0.7$ fm$^{-1}$.

The ${\Lambda}$-nuclear interaction parameters, ${\bar V}$, $V_{\sigma}$, $C_p$, and $W_o$ are
considered as unknown. These are then fitted as a function of the $B_{\Lambda}$ values that
have been calculated using the {\it s}-shell results. Taking these values of ${\bar V}$,
$V_{\sigma}$, $C_p$, and $W_o$ we again perform variational calculations that give us the
final results for ${_{\Lambda}^4}H$, ${_{\Lambda}^4}H^*$, and ${_{\Lambda}^5}He$. These results
are later used to analyze $_{\Lambda}^{17}O$ and  ${\Lambda}$ binding to nuclear matter.

\vskip 1.0 truecm
\centerline{\bf \S 3. Wave Function and Approach}
\vskip 12 truept

The trial variational wave function we adopt is of the following form:
$$
\mid \Psi_v \rangle = \Bigg [ {1 + \sum_{i<j<k} U_{ijk} + \sum_{i<j} U_{ij \Lambda}
+ \sum_{i<j} U_{ij}^{LS} + \sum_{i<j<k} U_{ijk}^{Tni}} \Bigg ]
\Bigg [ \prod_{i<j<k} f_{ijk}^c \Bigg ] \mid \Psi_p \rangle \,\,. \eqno(3.1)
$$
The pair wave function ${\mid \Psi_p \rangle}$ is a symmetrized product of two-body $(1+U_{ij})$ and $(1+U_{i \Lambda})$ correlation operators acting on a Jastrow trial function. This is written as
$$
\mid \Psi_p \rangle = \Biggl [ S \, {\prod_{i<j}^{A-1}} (1+U_{ij}) \Biggr ]
\Biggl [ S \, {\prod_{i=1}^{A-1}} (1+U_{i \Lambda}) \Biggr ] \mid \Psi_J \rangle \,\,. \eqno(3.2)
$$

$U_{ij}$ in Eq.(3.2) is defined as
$$
U_{ij} = \sum_{p=2,6} \Big [ \prod_{k \neq{i,j}} f_{ijk}^p (r_{ik},r_{jk})
\Big ] u_p(r_{ij}) \, O_{ij}^P \eqno(3.3)
$$

\noindent with $O_{ij}^P = [ {1, \vec \sigma_i \cdot \vec \sigma_j, S_{ij}} ] \otimes
[ {1, \vec \tau_i, \cdot \vec \tau_j} ]$. $S_{ij}$ is the tensor operator. ${\vec \sigma}$ and $\vec \tau$ are the spin and isospin operators, respectively. The factor $f_{ijk}^p$ suppresses spin-isospin correlations between two nucleons in the presence of a third one.

In principle, the $U_{i \Lambda}$ correlation will consist of a spin and a Majorana
space-exchange operator [16]:
$$
U_{i \Lambda} = \alpha_{\sigma} u_{\sigma}(r_{i \Lambda}) \, \vec \sigma_{\Lambda} \cdot
\vec \sigma_i + \alpha_{px} u_{px}(r_{i \Lambda}) P_x \,\,, \eqno(3.4)
$$
\noindent where $\alpha_{\sigma}$ and $\alpha_{px}$ are variational parameters and $P_x$ is
the space-exchange operator. In our calculations we have not included the space-exchange
correlations since the calculations become complicated and time consuming. In any case the
effect of these correlations is expected to be small for {\it s}-shell hypernuclei.

The spin dependent correlation $u_{\sigma}$ given in Eq.(3.4) is defined as
$$
u_{\sigma}= {(f_s^{\Lambda} - f_t^{\Lambda}) \over  {f_c^{\Lambda}}}  \,\,, \eqno(3.5)
$$

\noindent where $f_{c}^{\Lambda}$ is the spin-average correlation function. $f_s^{\Lambda}$
and $f_t^{\Lambda}$ are the solutions of the quenched ${\Lambda}${\it N} potential in singlet
and triplet states, respectively, which are given by the following relation:
$$
\Biggl [ -{\hbar^2 \over {2 \mu_{\Lambda N}}} \nabla^2 + {\tilde V}_{s/t} (r_{\Lambda N}) + V_{\Lambda N}^a \Biggr ] f_{s/t}^{\Lambda } = 0 \,\,, \eqno (3.6)
$$

\noindent where ${\tilde V}_{s/t}$ is the quenched ${\Lambda }${\it N} potential in
singlet/triplet state. $\mu_{\Lambda  N}$ is the reduced mass of the ${\Lambda}$-{\it N} pair, while $V_{\Lambda  N}^a$ is an auxiliary potential.

The {\it Jastrow} wave function ${ \mid \Psi_J \rangle}$ is given by
$$
\mid \Psi_J \rangle = \Big [ {\prod_{i=1}^{A-1}} f_c^{\Lambda}(r_{i \Lambda})
{\prod_{i<j}^{A-1}} f_c(r_{ij})  \Big ] \mid \Phi \rangle \,\,. \eqno(3.7)
$$
\noindent Here ${\mid \Phi \rangle}$ is an antisymmetric product of single particle wave
function with the desired ({\bf J},{\bf T}). The initial uncorrelated state $\Phi$ has no
coordinate dependence and is real. For example, consider the following $\Phi$ states for
${_{\Lambda}^4}H$ and ${_{\Lambda}^4}H$$^{*}$ expressed in the spin-isospin basis with the
appropriate ({\bf J},{\bf T}) states,

\hskip 1.0 truecm ${_{\Lambda}^4}H$ (J$=0$, T=${1 \over 2}$) $=$ {\bf C} $\biggl ( {1 \over 2},
{1 \over 2}, 0; {1 \over 2}, -{1 \over 2}, 0 \biggr ) \, \, {^3}$H$_{j={1 \over 2}}^{1 \over 2}
\, \, \Lambda_{1 \over 2}^{-{1 \over 2}}$ +
\par \hskip 4.7 truecm  {\bf C} $\biggl ( {1 \over 2}, {1 \over 2}, 0; -{1 \over 2},
{1 \over 2}, 0 \biggr ) \, \, {^3}$H$_{j={1 \over 2}}^{-{1 \over 2}} \, \, \Lambda_{1 \over 2}^{1
\over 2}$  \hskip 2.6 truecm (3.8)
\par \noindent and

{\baselineskip 20 truept
\hskip 0.85 truecm ${_{\Lambda}^4}H$$^{*}$ (J$=1$, T=${1 \over 2}$) $=$ {\bf C} $\biggl
( {1 \over 2}, {1 \over 2}, 1; {1 \over 2}, {1 \over 2}, 1 \biggr ) \, \, {^3}$H$_{j={1
\over 2}}^{1 \over 2} \, \, \Lambda_{1 \over 2}^{1 \over 2}$   $\,\,$, \hskip 2.6 truecm (3.9)

}
\vskip 0.5 truecm

\noindent where {\bf C} represents the {\it Clebsch Gordon}
coefficients[17]. ${^3}$H$_{j={1 \over 2}}^{1 \over 2}$ and
${^3}$H$_{j={1 \over 2}}^{-{1 \over 2}}$ are the uncorrelated
$\Phi$s for triton,  whereas $\Lambda_{1 \over 2}^{-{1 \over 2}}$
and $\Lambda_{1 \over 2}^{1 \over 2}$ are the spin-down and
spin-up states, respectively of the $\Lambda$ particle.

The spin-orbit correlation $U_{ij}^{LS}$ is given by
$$
U_{ij}^{LS}= \bigl [ u_{ls}(r_{ij}) + u_{ls \tau}(r_{ij}) \, {\vec \tau_i \cdot \vec \tau_j} \bigr ]
\, {({\bf L} \cdot {\bf S})}_{ij} \,\,. \eqno (3.10)
$$

The eight radial functions, $f_c(r_{ij})$, $u_{p=2,6}(r_{ij})$, $u_{ls}(r_{ij})$, and  $u_{ls \tau}(r_{ij})$ are obtained from approximate two-body Euler-Lagrange equations with variational parameters[18].

${U_{ijk}^{Tni}}$ is a three-body correlation induced by the three-nucleon interaction,
$V_{ijk}$. The other correlations incorporated in the wave function are a spatial three-body
{\it NNN} correlation $f_{ijk}^c$, along with $U_{ijk}$ that consists of a spin-orbit and an
isospin three-body correlation. Further details can be found in Ref. [19].

The three-body $\Lambda${\it NN} correlation $U_{ij \Lambda }$ has the following form:
$$
U_{ij \Lambda }= {\tilde V }_{ij \Lambda}(\hat \delta_1, \hat \delta_2) \,\,, \eqno (3.11)
$$
\noindent where ${\tilde V }_{ij \Lambda}$  differs from  ${ V }_{ij \Lambda}$ through the
cutoff factor $c$ of the usual Yukawa and tensor functions. $\hat \delta_1$ and $\hat
\delta_2$ are variational parameters which multiply the $\Lambda${\it NN} interaction parameters, $C_p$ and $W_o$, respectively.

No attempt has been made to vary the two-body {\it NN} and three-body {\it NNN} correlation
parameters[19] as their effect has been found to be small[20] and only the variational parameters of the wave function pertaining to ${\Lambda}$ have been varied to obtain a minimum in the energy. The optimum values of these parameters which are used in our final calculations are given in Tables[1] and [2].

We calculate energy expectation values using Monte Carlo (MC) integration[21,22]. The expectation values are sampled both in configuration space and in the order of operators in the wave function by following a Metropolis random walk[23]. The mathematical expressions used to evaluate the energy expectation  values are given below.

The energy expectation value for the pure nucleus is given by
$$
\langle E_N^{A-1} \rangle = {\langle \Psi_N^{A-1} \mid H_N^{A-1} \mid \Psi_N^{A-1}
\rangle \over \langle \Psi_N^{A-1} \mid \Psi_N^{A-1} \rangle} \,\,, \eqno(3.12)
$$
where $\Psi_N^{A-1}$ is the wave function of the mass (A-1) nucleus and $H_N^{A-1}$ is the nuclear Hamiltonian.

The energy expectation value for the hypernucleus is given by
$$
\langle E_H^{A} \rangle = {\langle \Psi_H^{A} \mid H_H^{A} \mid \Psi_H^{A} \rangle
\over \langle \Psi_H^{A} \mid \Psi_H^{A} \rangle} \,\,, \eqno(3.13)
$$
where $ \Psi_H^{A}$ is the wave function of the mass `A' hypernucleus and $H_H^{A}$ is the hypernuclear Hamiltonian.

Therefore, binding energy of $\Lambda$ to the hypernucleus is given by
$$
-B_{ \Lambda} = \langle E_H^{A} \rangle - \langle E_N^{A-1} \rangle \,\,. \eqno(3.14)
$$

The nuclear and hypernuclear wave functions, $\Psi_N^{A-1}$ and $\Psi_H^{A}$ are optimized with respect to the variational parameters to obtain the minimum in the energies.

The $B_{ \Lambda}$ value for each hypernuclei is calculated from the variational
results using Eq.(3.14). The $B_\Lambda$ value is thus written as a function of
the adjustable parameters in the $\Lambda$ Hamiltonian $H_{\Lambda}$, and is used to
determine the set of parameters which are consistent with the experimental $B_{\Lambda}$
values[26,27].

\vskip 1.0 truecm
\centerline{ \bf \S 4. Results and Discussion}
\vskip 12 truept
Table[3] gives the variational results for the nuclei, namely, ${^4}He$ and ${^3}H$ calculated
using the two-body {\it NN} Argonne $\upsilon_{18}$ interaction[5] and three-body
{\it NNN} Urbana $IX$ interaction[6,7] with relevant correlations. The numbers
appearing in parentheses in all the tables in this work indicate the statistical error in the
last digit. These calculations have been performed on similar lines as those by Wiringa
$\it{et \, al.}$[5,19] and the results conform to theirs. These results also check very
well with the recent calculations of Forest $\it{et \, al.}$[24], who use a truncated
version of the Argonne $\upsilon_{18}$ interaction.

Next we calculate the energy expectation values for the {\it s}-shell hypernuclei, namely
${_{\Lambda}^4}H$, ${_{\Lambda}^4}H$$^{*}$, and ${_{\Lambda}^5}He$ using the two-body {\it NN} Argonne $\upsilon_{18}$ and three-body {\it NNN} Urbana $IX$ interactions along with the two-body $\Lambda${\it N} and three-body $\Lambda${\it NN} interactions with appropriate correlations incorporated in the wave function. Variational calculations have been performed for different values of spin-average potential strength, $\bar V$ (6.10, 6.15, and 6.20 MeV). The different values of the space-exchange parameter $\epsilon$ used in this work are 0.24
($\bar V$=6.20 MeV), 0.19 ($\bar V$=6.15 MeV), and 0.14 ($\bar V$=6.10 MeV)
[25].

The bulk calculations consist of the energy expectation values for each $\bar V$ as a function of the interaction parameters, $V_{\sigma}$, $C_p$, and $W_o$. In Table[4] we
illustrate one such set of results for the ground state of
${_{\Lambda}^4}H$. Our results demonstrate that the $B_{\Lambda}$
values for ${_{\Lambda}^5}He$, ${_{\Lambda}^4}H$, and
${_{\Lambda}^4}H$$^{*}$ show similar trends with the spin-average
potential strength $\bar V$ and ${\Lambda}${\it NN} interaction
parameters, $C_p$ and $W_o$. As expected, $B_{\Lambda}$ increases
with $\bar V$. $B_{\Lambda}$ also increases significantly with the
increase in $C_p$, while it decreases with $W_o$. As expected, the
dependence on $\bar V$, $C_p$, and $W_o$ is more pronounced for
$A=5$ than for $A=4$ systems. This result is in accord with the
earlier calculations where only simplified {\it NN} interactions
have been used[4].

An important goal of the present study is to learn about the role of $V_\sigma$ , $C_p$,
and $W_o$ through the $0^+- 1^+$ energy splitting in ${_{\Lambda}^4}H$ and ${_{\Lambda}^4}
H$$^{*}$. We place limits on the values of these parameters, consistent with the following
experimental $B_{\Lambda}$ values:
$$
B_{\Lambda} ({_{\Lambda}^4}H) = 2.22 \pm 0.04 MeV,
$$
$$
B_{\Lambda} ({_{\Lambda}^4}H^{*}) = 1.12 \pm 0.06 MeV,
$$
$$
B_{\Lambda} ({_{\Lambda}^5}He) = 3.12 \pm 0.02 MeV.    \eqno (4.1)
$$
\noindent Values of $B_{\Lambda}$ $({_{\Lambda}^4}H$) and $B_{\Lambda}$ $({_{\Lambda}^4}H$$^{*}$)
 are averages for those of ${_{\Lambda}^4}H$ and ${_{\Lambda}^4}He$.
Limits on the parameters $V_\sigma$ , $C_p$, and $W_o$  are
determined by the uncertainties in the experimental $B_{\Lambda}$
values.

For a given value of $\bar V$ we did a $\chi^2$ fit for the calculated energy expectation values
according to the relation
$$
B_{\Lambda}(V_{\sigma}, W_o, C_p)= y_1 \, V_{\sigma} + y_2 \, W_o + y_3 \, C_p + y_4 \, W_o^2
+ y_5 \, C_p^2 + y_6 \, W_o C_p +  B_{\Lambda}^{o} \,\,, \eqno(4.2)
$$

\noindent where $B_{\Lambda}^{o}$ is the corresponding value of $B_{\Lambda}$ for $V_{\sigma}
=C_p=W_o=0$ for each hypernuclear species. The coefficients $y_{1-6}$ are varied to give a
minimum in the $\chi^2$ that is defined as
$$
\chi^2 = {1 \over N} \sum^{N}_{i=1} \Biggl[ {{{B_\Lambda {(V_\sigma, C_p, W_o)} - B_\Lambda} \over {{\Delta B_\Lambda}}} \Biggr]}^2_i,  \, \eqno(4.3)
$$

Here $N$ is the total number of energy calculations for a particular hypernucleus with
different values of $V_\sigma$, $C_p$, and $W_o$. $B_\Lambda$ is the calculated value of the
$\Lambda$ separation energy and $\Delta{B_\Lambda}$ is the corresponding Monte Carlo
statistical error. The values of the coefficients $y_{1-6}$, as determined from this procedure,
are displayed in Table[5].  In all the cases considered the $\chi^2$ values are $\le$ 1 which
demonstrates the goodness of the fit. There shall be correlated error bars on the coefficients
$y_{1-6}$ which would be reflected in the uncertainties in determining $V_\sigma$, $C_p$, and
$W_o$. We find it more convenient to consider the uncertainties in the experimental values
while placing limits on $V_\sigma$, $C_p$, and $W_o$. We hope to compensate some of the
uncertainties associated with the $y$'s by giving a generous allowance to the experimental
$\Delta B_\Lambda$ values as well as by taking into account the Monte Carlo statistical errors
in the calculation of the energies.

We use the coefficients $y_{1-6}$ of Table[5] to obtain a fit with
respect to the experimental $B_\Lambda$ values, treating
$V_\sigma$, $C_p$, and $W_o$ as parameters to determine the best
fit. We again construct a $\chi^2$ fit using Eq.(4.3) but now ``$N$''
refers to the factor ``3'' for the three hypernuclear species and
$B_\Lambda$ refers to the experimental $B_\Lambda$ values. The
$\chi^2$ fit is minimized with respect to $C_p$ and $W_o$ for a
given value of $V_\sigma$. In Fig. 1 we plot $\chi^2$, $C_p$,
and $W_o$ as a function of $V_\sigma$ for $\bar V$ = 6.15 MeV. It is seen that both $C_p$ and $W_o$ decrease with increase
in $V_\sigma$, the effect being more pronounced for $C_p$. Fig.
2 displays the calculated  values of $B_\Lambda$ as a function of
$V_\sigma$  for ${_{\Lambda}^4}H$, ${_{\Lambda}^4}H$$^{*}$, and
${_\Lambda^5}He$. Within the accuracy of the graphs the
$B_\Lambda$ values for ${_{\Lambda}^4}H$ and ${_\Lambda^5}He$ do
not show any dependence on $V_\sigma$ in the range 0.09$-$0.26
MeV. As one may expect, the $B_\Lambda$ values for
${_{\Lambda}^4}H$$^{*}$ depend sensitively on $V_\sigma$, and thus
in turn on the spin dependence of $C_p$ and $W_o$.

The $\chi^2$ stays very close to zero (which corresponds to almost
an exact fit to $B_\Lambda$(exp) values) for $V_\sigma$ =
0.176$\pm$0.015 MeV, $C_p$ = 1.64$\mp$.03 MeV and
$W_o$ = 0.026$\mp$0.001 MeV.  Most of the deviation from
zero of the $\chi^2$ values in Fig. 1 arise from
${_{\Lambda}^4}H$$^{*}$. The dotted horizontal lines of Fig. 2
display the limits on the experimental $B_\Lambda$ value of
${_{\Lambda}^4}H$$^{*}$, consistent with the experimental error
bar of $\pm$ 0.06 MeV. This places limits on the values of
$V_\sigma$, thus, in turn on $C_p$ and $W_o$. However, the actual
error bars on these parameters would be larger due to Monte Carlo
statistical errors. To take this into account, we made a number of
energy calculations for $V_\sigma$ in the range 0.10 to 0.24 MeV. The corresponding values of $C_p$ and $W_o$ have been taken
from the calculations of Fig. 1. We could obtain acceptable fits
to the energies for $V_\sigma$ in the range 0.12$-$0.23 MeV. This gives for $\bar V$=6.15 MeV:
$$
V_\sigma = 0.176 \pm 0.05, C_p = 1.64 \mp 0.15, W_o = 0.026 \mp
0.003.              \eqno (4.4)
$$

A similar study for $\bar V$=6.20 MeV gives
$$
V_\sigma = 0.125 \pm 0.05, C_p= 1.52 \mp 0.15, W_o = 0.025 \mp 0.003.
 \eqno(4.5)
$$

We have been able to obtain good fits only for $\bar V$ = 6.15 and
6.20 MeV. The value of $\bar V$ = 6.10 MeV does not
reproduce the correct binding energies of {\it s}-shell
hypernuclei. Therefore, we have not carried out any error analysis
for $\bar V$=6.10 MeV. For the sake of completeness we
mention the best parameter values for $\bar V$=6.10 MeV:

$$
V_\sigma = 0.193, C_p= 1.84, W_o = 0.027.   \eqno (4.6)
$$

We can note from Table[6(A)] that the $\Lambda${\it N} spin
potential has a non zero contribution even in a closed-shell
system such as ${_{\Lambda}^5}He$. This arises because of the
$\Lambda${\it N} spin-spin correlations incorporated in the wave
function.

Comparing the results for the core nuclei (Table[3]) with the
results for the hypernuclei (Tables[6(A-C)]) we note, in general,
a shrinking of the core nuclei by about $20\%$ in all the
hypernuclei due to the presence of the ${\Lambda}$ particle. This
decrease in radii of the core nuclei would imply that the lambda
wave functions are closer for larger $A$. This also contributes to
the fact that the dependence of $B_\Lambda$ on $\bar V$, $C_p$,
and $W_{o}$ is more pronounced for the mass 5 than for the mass 4
hypernuclei. The change in the $d$-state probability is found to be
small in all cases.

Table[7] gives the breakup of the $0^{+}-1^{+}$ splitting
contributions in ${_{\Lambda}^4}H$ and ${_{\Lambda}^4}H$$^{*}$
arising from the $\Lambda${\it N} spin-dependent strength
$V_{\sigma}$ and the three-body $\Lambda${\it NN} interaction
$V_{\Lambda NN}$ for different values of spin-average strength
$\bar V$. It can be noted from this table that the energy
difference between ${_{\Lambda}^4}H$ and ${_{\Lambda}^4}H$$^{*}$
is consistent with the total contribution from $V_{\sigma}$ and
$V_{\Lambda NN}$ within the error bars of the MC
calculations. It can be seen that a large part ($\sim {2 \over
3}$) of the splitting comes from the three-body $\Lambda${\it NN}
potential. The two-body contribution arising from $V_{\sigma}$ is
around $\sim {1 \over 3}$ of the total splitting. This is in
contrast to the earlier studies [4,28,29] wherein the
$0^{+}-1^{+}$ splitting has been thought to have arisen mainly
from the spin dependence of the two-body $\Lambda${\it N}
potential. The present study clearly demonstrates that $V_{\Lambda
NN}$ plays a significant role in explaining the splitting. This
also results in a reduced $V_{\sigma}$ as compared to the value of
0.23$\pm$0.02 found in Ref. [4], though
in our case the error bar on $V_\sigma$ is much larger due to
reasons discussed earlier. In the present study for
$V_\sigma$=0.23, half of the splitting arises because of
$V_\sigma$ and the remaining half from the three-body $\Lambda NN$
forces. For the extreme case, in particular, for
$V_{\sigma}$=0.12, the three-body forces contribute nearly $3
\over 4$ of the total splitting. It would be desirable to have an
independent fix on $V_{\sigma}$, for example, from a more refined
$\Lambda p$ scattering data. This can enlighten us further on the
three-body $\Lambda${\it NN} forces.

The Majorana space-exchange contribution for the various
hypernuclei have been found to be small but significant in all
{\it{s}}-shell hypernuclei. It is in the range $0.1-0.7$ MeV
for $\epsilon$ in the range $0.14-0.24$ and as expected has a
linear dependence with the Majorana exchange parameter $\epsilon$.

A few variational calculations have been carried out with only the
${\Lambda}${\it NN} potentials and no ${\Lambda}${\it NN}
correlations. We find that without the ${\Lambda}${\it NN}
correlations ${_{\Lambda}^5}He$ is not bound. We also notice that
the contributions from $V^{2 \pi}_{\Lambda NN}$ and
$V^{DS}_{\Lambda NN}$ become more repulsive without the
correlations and the total contribution from the three-body
${\Lambda}${\it NN} potentials become positive thereby decreasing
$B_{\Lambda}$. We have performed a few energy calculations for
${_{\Lambda}^5}He$, ${_{\Lambda}^4}H$, and ${_{\Lambda}^4}H$$^{*}$
but with the three-body ${\Lambda}${\it NN} part of the
Hamiltonian completely switched off . We find in this case that
${_{\Lambda}^5}He$ is overbound by about 2.34 MeV for $\bar
V=6.20$ MeV and by about 1.39 MeV for $\bar V=6.15$
MeV. In general, the results for ${_{\Lambda}^4}H$ show that
it is underbound while ${_{\Lambda}^4}H$$^{*}$ is overbound
without the three-body ${\Lambda}${\it NN} interactions for all
values of $\bar V$. Both these studies show the importance of the
three-body ${\Lambda}${\it NN} potentials and correlations in
obtaining a consistent fit to the $B_\Lambda$ values for all the
{\it s}-shell hypernuclei considered here. In particular,
we notice the importance of the $\Lambda${\it NN} correlations
$f^{2 \pi}_{\Lambda NN}$ on the effect of TPE $\Lambda${\it
NN} forces $V^{2 \pi}_{\Lambda NN}$. These correlations reduce
reasonably the repulsive three-body contribution to an attractive
contribution implying a strong non-linear dependence on $C_p$.

\vskip 0.5 truecm

\noindent {\it Implications of {\it s}-shell results on
$_{\Lambda}^{17}O$}: We now examine the $_{\Lambda}^{17}O$
hypernucleus in relation to the two- and three-body
$\Lambda$-nuclear potential parameters that we find from our
analysis of {\it s}-shell hypernuclei as described earlier.
Usmani, Pieper, and Usmani$^{16)}$ (referred to as UPU) have
carried out MC calculations for the $_{\Lambda}^{17}O$
hypernucleus using the $\upsilon_6$ part of the older Argonne
$\upsilon_14$ potential [9]. For the three-nucleon potential they
use the same form (Urbana model) as in the present study but with
different strength parameters ($A_o$=$-$0.0333 and $U_o$=0.0038).
Their trial wave function consists of pair and triplet operators
acting on a single-particle determinant. In many respects, these
calculations are similar to our present calculations. The
difference lies in the treatment of non-central correlations for
which they use the Cluster Monte Carlo method with up to
four-baryon clusters. The central correlations are treated
exactly. We carry out this study in the hope of analyzing further
our estimated {\it s}-shell ${\Lambda}$-nuclear parameters. UPU
have given the following empirical relation for $C_p$ in the range
0-1 MeV and $W_o$ in the range 0-0.02 MeV:
$$
B_{\Lambda}= 27.3 -8.9 \,C_p + 11.2 \, C_p^2 + 870.0 \, W_o . \eqno (4.7)
$$

This equation relates the $B_\Lambda$ of $_{\Lambda}^{17}O$ with
$C_p$ and $W_o$. In order to test the consistency of our results
with $B_\Lambda$($_\Lambda^{17}O$), we assume that relation (4.7)
holds for our values of $C_p$ and $W_o$. UPU have done
calculations for spin-average strength $\bar V$=6.16 MeV
with space-exchange parameter $\epsilon$= 0.3. Our values of $\bar
V$=6.15 MeV with $\epsilon$=0.17 are closest to theirs. We
thus need to modify (4.7) for our values of $\bar V$ and
$\epsilon$. Unfortunately, there is no simple method to scale
relation (4.7) for $\bar V$=6.15 MeV, as the scaling can be
considerably non-linear. However, since the two values of $\bar V$
are very close, we assume that this will not affect the results
much. The correction for $\epsilon$ is simple, since in the
absence of space-exchange correlation the space-exchange energies
are expected to be linear with $\epsilon$. Thus, relation (4.7)
can be modified as:
$$
B_{\Lambda}= 27.3 + (\epsilon -\epsilon_o) \langle \, (1-P_x) \,
V_{\Lambda N} \rangle -8.9 \,C_p + 11.2 \, C_p^2 + 870.0 \, W_o
\,\,, \eqno (4.8)
$$
\noindent where $\epsilon_o$=0.3 as taken by UPU, $P_x$ is the
space-exchange operator and $\langle$$V_{\Lambda N}\rangle$ is the
energy expectation value of the ${\Lambda}${\it N} potential.

Using the entries for $\upsilon_o(r)(1-\epsilon)$ and
$\upsilon_o(r) \, \epsilon P_x$ from Table II of Ref. [16] and
using our values of $C_p=1.6407$ and $W_o=0.0255$ for $\bar
V=6.15$ MeV we obtain

\hskip 4.0 truecm B${_\Lambda}$ = 23.3 $\pm$ 1.6 MeV $\,\,$.

This is considerably larger than the empirical estimate$^{16)}$
$\sim$ 13.0 $\pm$ 0.4 MeV. Thus the results of the present
study are incompatible with those of Ref. [16]. The reason for
this incompatibility may largely lie in the use of $\upsilon_6$
part of the $\upsilon_{18}$ hamiltonian for $_{\Lambda}^{17}O$.
Another reason can be attributed to the use of relations (4.7) and
(4.8) for large values of $C_p$ and $W_o$ for which these
relations may not be adequate. This discrepancy can probably be
resolved by carrying out calculations for $_{\Lambda}^{17}O$ with
Argonne $\upsilon_{18}$ Hamiltonian, which, at the moment is an
extremely challenging task.

\vskip 0.5 truecm

\noindent {\it Implications of s-shell results on
${\Lambda}$ binding to nuclear matter}: The presence of a
$\Lambda$ particle inside nuclear matter can reveal information on
the $\Lambda$-nuclear interactions. The well depth {\it D} is
identified with the separation energy for a ${\Lambda}$ in nuclear
matter. It is an important parameter which can help to distinguish
between different ${\Lambda}${\it N} potentials and also throw
light on the ${\Lambda}${\it NN} interaction. ${\Lambda}$ binding
to nuclear matter can put further constraints on the potential
parameters, namely $\bar V$, $V_{\sigma}$, $W_o$, and $C_p$. With
this aim in mind we have performed calculations for {\it D} using
the {\it Fermi-Hypernetted-Chain (FHNC)} technique[4] to
calculate the energy expectation values.

We have calculated the well depth {\it D} variationally using the
same underlying principle as for our {\it s}-shell hypernuclei.
Our discussion on {\it D} is based on the results given in
Table[8]. The empirical value of {\it D} is now fairly
well established at 29 $\pm$ 1 MeV  [30,31] at the
normal nuclear matter density of $\rho_{o} \approx 0.16 \, {\rm
fm}^{-3}$. These results clearly indicate that $\Lambda$ is
underbound at the normal nuclear matter density, $\rho_{o} \approx
0.16 \, {\rm fm}^{-3}$. This indeed is a disturbing feature.
Bodmer and Usmani[4] have found that it is possible to obtain
a consistent phenomenology with hypernuclear interactions which
include the {\it s}-shell and the medium and heavy hypernuclei as
well as $\Lambda$ binding to nuclear matter. Our results indicate
that with the present available techniques of treating the nuclear
matter this is not possible. The resolution of this paradox
perhaps lies in the proper handling of the three-body
correlations, particularly the $\Lambda${\it NN} correlations for
nuclear matter. It may be noted that the contribution from the
TPE $\Lambda${\it NN} forces $V^{2 \pi}_{\Lambda NN}$ for
nuclear matter is always positive[4]. On the other hand, $V^{2
\pi}_{\Lambda NN}$ for {\it s}-shell hypernuclei and
${_{\Lambda}^{17}}O$ is always negative and substantial. The
three-body $\Lambda${\it NN} correlations that are taken in
nuclear matter calculations always pertain to the {\it
s}shell$^{4)}$. The reason for adopting this correlation lies in
its simplicity. At present the techniques for incorporating the
realistic three-body $\Lambda${\it NN} correlations for nuclear
matter are not sufficiently developed as compared to those for the
{\it s}-shell hypernuclei incorporated in this work. These affect
the contribution from $V^{2 \pi}_{\Lambda NN}$ quite
substantially, even to the extent of reversing its sign in the
presence of the $\Lambda${\it NN} correlations as can be seen in
the present as well as in the ${_{\Lambda}^{17}}O$ studies. The
correct incorporation of the three-body correlations in nuclear
matter may affect the results to quite an extent, particularly
those at high densities[32]. The incorporation of these
correlations is indeed a challenging task and is very much needed
for the present work as well as for other related studies.

\vskip 1.0 truecm
\centerline{ \bf \S 5. Conclusion and Comments}
\vskip 12 truept

From the results discussed in the previous section we note that
the values of the spin-average strength $ \bar V$= 6.20 and 6.15
MeV give a reasonable description of the {\it s}-shell
hypernuclei. We have been able to provide a consistent account of
the $B_{\Lambda}$ values of  ${_{\Lambda}^4}H$,
${_{\Lambda}^4}H$$^{*}$, and ${_{\Lambda}^5}He$ using the
realistic Argonne $\upsilon_{18}$ {\it NN} interaction and
Urbana IX {\it NNN} interaction alongwith
${\Lambda}$-nuclear interactions with appropriate correlations.
Our results for $B_{\Lambda}$ show very similar trends with the
spin-average strength $\bar V$ of the ${ \Lambda}${\it N}
interaction, and the ${ \Lambda}${\it NN} interaction parameters,
$C_p$ and $W_o$ for all the {\it{s}}-shell hypernuclei considered.

An important conclusion of our study is that about 25\% to 50\% of
the $0^{+}-1^{+}$ splitting energy between the ${_{\Lambda}^4}H$
and ${_{\Lambda}^4}H$$^{*}$ comes from the $\Lambda${\it N}
spin-dependent strength $V_{\sigma}$. The earlier
studies [4,28,29] attribute a larger part of the splitting to
the spin dependence of the two-body $\Lambda${\it N} interactions.
In contrast, our study indicates that the major part ($\sim$ 50\%
to 75\%) of the splitting is generated by the three-body
$\Lambda${\it NN} forces.

Our study on ${\Lambda}$-binding to nuclear matter shows that
${\Lambda}$ is underbound. This indicates the fact that there is a
need to include the three-body correlations while treating nuclear
matter. This would require a different technique altogether and is
a challenging problem in itself. Our analysis on
$_{\Lambda}^{17}O$ also indicate the importance of the non-central
correlations. It is possible that the inconsistency between the
results of Ref. [16] and our {\it s}-shell results is due to
their neglecting terms with higher than four-baryon clusters and
thereby neglecting the contributions from the non-central
clusters. Moreover, our values of the $\Lambda${\it NN}
interaction parameters, $C_p$ and $W_o$ are higher than those of
Ref. [16] and which, in turn, would induce stronger
$\Lambda${\it NN}  correlations.

Contrary to the findings by Bando $\it{et \, al.}$ [33] and
Shinmura $\it{et \, al.}$ [34] regarding the effect of tensor
forces on the overbinding problem, we find that the tensor forces
do not play a significant role. Further, separate studies by Carlson[28]
and Hiyama $\it{et \, al.}$ [29] and  on four- and
five-body hypernuclei have shown that the binding energies and the
splitting energies are not reproduced correctly using the Nijmegen
interactions which have strong tensor terms. However, the small
suppression effects expected from $\Lambda${\it N} tensor forces
are already implicitly included in our phenomenological dispersive
$\Lambda${\it NN} force. Moreover, the Argonne
$\upsilon_{18}$ potential used in our study has a weak tensor part
and this in fact, provides a much better binding to nuclei$^{5)}$.

\vskip 1.0 truecm

\centerline{ \bf ACKNOWLEDGMENTS}
\vskip 12 truept

We would like to thank Dr. R. B. Wiringa for providing his
computer routines for the mass 3 and mass 4 nuclei. We gratefully
acknowledge Professor A. R. Bodmer for fruitful discussions and
suggestions during the course of this work.

\vskip 1.0 truecm
\centerline{\bf REFERENCES}
\vskip 12 truept

{\baselineskip= 20 truept

[1] Dean Halderson, Phys. Rev. {\bf {C61}}, 034001(2000)

[2] Y. Yamamoto and H. Bando, Prog. Theo. Phys. Supp. {\bf {81}}, 9 (1985)

[3] A. R. Bodmer, Q. N. Usmani, and J. Carlson, Phys. Rev. {\bf{C29}}, 684 (1984)

[4] A. R. Bodmer and Q. N. Usmani, Nucl. Phys. {\bf{A477}}, 621 (1988)

[5] R. B. Wiringa, V. G. J. Stoks, and R. Schiavilla, Phys. Rev. {\bf C51}, 38 \par {\hskip 1.2 em} (1995)

[6] J. Carlson, V. R. Pandharipande, and R. B. Wiringa, Nucl. Phys. {\bf A401}, \par {\hskip 1.2 em} 59 (1983)

[7] B. S. Pudliner, V. R. Pandharipande, J. Carlson, and R. B. Wiringa,
\par {\hskip 1.2 em} Phys. Rev. Lett. {\bf 74}, 4396 (1995)

[8] Q. N. Usmani, Nucl. Phys. {\bf A340}, 397 (1980)

[9] R. B. Wiringa, R. A. Smith, and T. L. Ainsworth, Phys. Rev. {\bf C29}, 1207 \par {\hskip 1.2 em} (1984)

[10] J. R. Bergervoet, P. C. van Campen, R. A. M. Klomp, J. L. de Kok, T. \par {\hskip 1.6 em}
A. Rijken, V. G. J. Stoks, and J. J. de Swart, Phys. Rev. {\bf C41}, 1435 \par {\hskip 1.6 em} (1990)

[11] V. G. J. Stoks, R. A. M. Klomp, M. C. M. Rentmeester, and J. J. de \par {\hskip 1.6 em}
Swart, Phys. Rev. {\bf C48}, 792 (1993)

[12] J. Fujita and H. Miyazawa, Prog. Theor. Phys. {\bf 17}, 360 (1957)

[13] I. E. Lagaris and V. R. Pandharipande, Nucl. Phys. {\bf A359}, 331 (1981)

[14] Thomas Schimert, D. J. Stubeda, M. Lemere, and Y. C. Tang, Nucl. \par {\hskip 1.6 em} Phys. $\bf{A343}$, 429 (1980); Wang Xi-chang, H. Takaki, and H. Bando, \par {\hskip 1.6 em} Prog. Theor. Phys. $\bf{76}$, 865 (1986)

[15] R. K. Bhaduri, B. A. Loiseau, and Y. Nogami, Annals of Phys. {\bf 44}, 57 \par {\hskip 1.6 em} (1967)

[16] A. A. Usmani, S. C. Pieper, and Q. N. Usmani, Phys. Rev. {\bf C51}, 2347 \par {\hskip 1.6 em} (1995)

[17] M. K. Pal, {\it Theory of Nuclear Structure} 
(Affiliated \par {\hskip 1.6 em} East-West, New Delhi) Appendix A, pg {\bf 606}

[18] R. B. Wiringa, Phys. Rev. {\bf{C43}}, 1585 (1991)

[19] A. Arriaga, V. R. Pandharipande, and R. B. Wiringa, Phys. Rev. {\bf C52},
\par {\hskip 1.6 em} 2362  (1995)

[20] S. Murali, Ph.D. Thesis, Jamia Millia Islamia, New Delhi, India (1995)

[21] J. Lomnitz-Adler, V. R. Pandharipande, and R. A. Smith, Nucl. Phys. \par {\hskip 1.6 em} {\bf A361}, 399 (1981)

[22] J. Carlson and R. B. Wiringa, {\it in Computational Nuclear Physics,} ed. by \par {\hskip 1.6 em} S. E. Koonin, K. Langanke, J. A. Maruhn, and M. R. Zirnbauer (Springer- \par {\hskip 1.6 em} Verlag, Berlin, 1991)

[23] N. Metropolis, A. W. Rosenbluth, M. N. Rosenbluth, A. H. Teller, and
\par {\hskip 1.6 em} E. Teller, J. Chem. Phys. {\bf{21}}, 1087 (1953)

[24] J. L. Forest, V. R. Pandharipande, and A. Arriaga, Phys. Rev. {\bf C60}, \par {\hskip 1.6 em} 014002 (1999)

[25] Q. N. Usmani and A. R. Bodmer (private communication)

[26] J. Pniewski and D. Zieminska, {\it in} Proceedings of the Conference on Kaon-\par {\hskip 1.6 em} nuclear interactions and Hypernuclei, Zvenigorod, pg {\bf 33} (1977)

[27] M. Bedjidian, E. Descroix, J. Y. Grossiord, A. Guichard, M. Gusakow, \par {\hskip 1.6 em} M. Jacquin, M. J. Kudla, H. Piekarz, J. Piekarz, J. R. Pizzi, and J. \par {\hskip 1.6 em} Pniewski, Phys. Lett. {\bf 83B}, 252 (1979), and references therein.

[28] J. Carlson, in {\it LAMPF Workshop on ($\pi$, K) Physics}, ed. B. F. Gibson, \par {\hskip 1.6 em} W. R. Gibbs, and M. B. Johnson, AIP Conf. Proc. No.{\bf 224} (AIP, New {\par {\hskip 1.6 em}} York, 1991)

[29] E. Hiyama, M. Kamimura, T. Motoba, T. Yamada, and Y. Yamamoto,
\par {\hskip 1.6 em} Nucl. Phys. {\bf A639}, 169c (1998)

[30] Q. N. Usmani and A. R. Bodmer, Nucl. Phys {\bf{A639}}, 147c (1998)

[31] Q. N. Usmani and A. R. Bodmer, Phys. Rev. {\bf C60}
055215 (1999)

[32] A. Akmal, V. R. Pandharipande, D. G. Ravenhall, Phys. Rev. {\bf C58}
1804 \par {\hskip 1.6 em} (1998) and references therein.

[33] H. Bando and I. Shimodaya, Prog. Theor. Phys. {\bf{63}}, 1812 (1986)

[34] S. Shinmura, Y. Akaishi, and H. Tanaka, Prog. Theor. Phys. {\bf 65}, 1290 \par {\hskip 1.6 em} (1981)

}

\vfill\break

$$ \vbox { \offinterlineskip  \vskip6pt
\def\qq{\hskip 0.3em}

\def\titlestrut{\vrule depth4pt height10pt width0pt}
\def\upstrut{\vrule height10pt width0pt}
\def\downstrut{\vrule height12pt width0pt}
\halign {\qq#\hfil &&\qq# &\qq\hfil#\hfil \cr
\multispan{16} \hfil {\bf TABLE 1. ${\Lambda}N$ Correlations Parameters} \hfil \titlestrut &\cr
\noalign {\vskip 0.4 em }
\noalign {\hrule depth0.005em}
\noalign {\vskip 0.08 em }
\noalign {\hrule depth0.005em}
\hskip2.0 em ${_{\Lambda}^A}Z$ && $\bar V$ && $\kappa_{\Lambda N}$ && $a_{\Lambda N}$ && $C_{\Lambda N}$ && $R_{\Lambda N}$ && $\alpha_s$ && $\alpha_{\sigma}$${^{ \dagger }}$ \upstrut &\cr
\noalign {\vskip 0.4 em }
\noalign {\hrule depth0.005em}
\noalign {\vskip 0.08 em }
\noalign {\hrule depth0.005em}
\hskip2.0em ${_{\Lambda}^5}He$ && 6.20&& 0.117&& 0.50 && 2.0 && 1.0 && 0.965 && 0.80 \upstrut &\cr
\hskip0.2em && 6.15&& \hskip2pt 0.110&& 0.50 && 2.0 && 1.0 && 0.970 &&  0.95 \upstrut &\cr
\hskip0.2em && 6.10&& \hskip2pt 0.095&& 0.50 && 2.0 && 1.0 && 0.940 &&  0.95 \upstrut &\cr
\noalign {\vskip 0.6 em }
\hskip2.0em ${_{\Lambda}^4}H$ && 6.20&& 0.12&& 0.70 && 2.0 && 1.0 && 0.95 && 0.70 \upstrut &\cr
\hskip0.2em && 6.15&& \hskip2pt 0.10&& 0.70 && 2.0 && 1.0 && 0.95 &&  1.20 \upstrut &\cr
\hskip0.2em && 6.10&& \hskip2pt 0.08&& 0.70 && 2.0 && 1.0 && 0.95 &&  0.70 \upstrut &\cr
\noalign {\vskip 0.6 em }
\hskip2.0 em ${_{\Lambda}^4}H$$^{*}$ && 6.20&& 0.095&& 0.70&& 2.0&& 1.0&& 1.0&& 0.70 \upstrut &\cr
\hskip0.2em && 6.15&& \hskip2pt 0.065&& 0.70 && 2.0 && 1.0 && 1.0 && 0.70 \upstrut &\cr
\hskip0.2em && 6.10&& \hskip2pt 0.050&& 0.70 && 2.0 && 1.0 && 1.0 &&  0.70 \upstrut &\cr
\noalign {\vskip 0.4 em }
\noalign {\hrule depth0.005em}
\noalign {\vskip 0.08 em }
\noalign {\hrule depth0.005em}
\noalign {\vskip 4pt}
\omit \quad {\srfont $\dagger$ Eq.(3.8); For all other correlation parameters refer to 16)
} \hidewidth \cr
}} $$

$$ \vbox { \offinterlineskip  \vskip6pt
\def\qq{\hskip 0.3em}

\def\titlestrut{\vrule depth4pt height10pt width0pt}
\def\upstrut{\vrule height10pt width0pt}
\def\downstrut{\vrule height12pt width0pt}
\halign {\qq#\hfil &&\qq# &\qq\hfil#\hfil \cr
\multispan{24} \hfil {\bf TABLE 2. $\Lambda${\it NN} Correlation Parameters for ${_{\Lambda}^5}He$, ${_{\Lambda}^4}H$, and ${_{\Lambda}^4}H$$^{*}$} \hfil \titlestrut &\cr
\noalign {\vskip 0.4 em }
\noalign {\hrule depth0.005em}
\noalign {\vskip 0.1 em }
\noalign {\hrule depth0.005em}
\hskip0.6 em Parameters && \hskip 1.2em $\bar V$=6.20 MeV && \hskip 0.8em $\bar V$=6.15 MeV && \hskip 0.8em $\bar V$=6.10 MeV  \upstrut &\cr
\noalign {\vskip 0.4 em }
\noalign {\hrule depth0.005em}
\noalign {\vskip 0.1 em }
\noalign {\hrule depth0.005em}
\noalign {\vskip 0.08 em }
\hskip2.0em $\hat \delta_1$${^{\dagger}}$ && \hskip 1.2em 0.364104&& 0.311733 && 0.257894 \upstrut &\cr
\noalign {\vskip 0.08 em }
\hskip2.0em $\hat \delta_2$${^{ \dagger}}$ && \hskip 1.2em 0.006096&& 0.004845 && 0.003766 \upstrut &\cr
\noalign {\vskip 0.4 em }
\noalign {\hrule depth0.005em}
\noalign {\vskip 0.1 em }
\noalign {\hrule depth0.005em}
\noalign {\vskip 4pt}
\omit \quad {\srfont $\dagger$ Eq.(3.11)
} \hidewidth \cr
}} $$

$$ \vbox { \offinterlineskip  \vskip6pt
\def\qq{\hskip 0.3em}

\def\titlestrut{\vrule depth4pt height10pt width0pt}
\def\upstrut{\vrule height10pt width0pt}
\def\downstrut{\vrule height12pt width0pt}
\halign {\qq#\hfil &&\qq# &\qq\hfil#\hfil \cr
\multispan{10} \hfil {\bf TABLE 3.} {\bf Variational results for $^3H$ and $^4He$} \hfil \titlestrut &\cr
\noalign {\vskip 0.5 em }
\noalign{\hrule depth0.005em}
\noalign{\vskip 0.08 em }
\noalign{\hrule depth0.005em}
\hskip 1.5 em Components && \hskip 0.2 truecm $^3H$ && \hskip 0.4 truecm $^4He$ \upstrut &\cr
\noalign {\vskip 0.4 em }
\noalign{\hrule depth0.005em}
\noalign{\vskip 0.08 em }
\noalign{\hrule depth0.005em}
\noalign {\vskip 0.2 em }
\hskip 0.8 em Kinetic energy&& \hskip0.5 truecm 50.76(4)&& \hskip0.8 truecm 106.85(6) \upstrut &\cr
\hskip 0.8 em {\it NN} Potential energy&& \hskip0.35 truecm -57.95(4)&& \hskip0.7 truecm -129.30(6) \upstrut &\cr
\hskip 0.8 em {\it NNN} Potential energy&& \hskip0.6 truecm -1.13(3)&& \hskip1.1 truecm -5.27(7) \upstrut &\cr
\hskip 0.8 em {\bf Total energy}&& \hskip0.35 truecm {\bf -8.32(2)}&& \hskip0.65 truecm {\bf -27.71(6)} \upstrut &\cr
\noalign {\vskip 0.2 em }
\hskip 0.8 em r.m.s.(proton)&& \hskip0.5 truecm 1.585(3)&& \hskip1.05 truecm 1.478(2) \upstrut &\cr
\hskip 0.8 em r.m.s.(neutron)&& \hskip0.5 truecm 1.731(4)&& \hskip1.05 truecm 1.478(2)
\upstrut &\cr
\hskip 0.8 em d-state probability&& \hskip0.3 truecm 0.0933(1)&& \hskip0.8 truecm 0.1512(2) \upstrut &\cr
\noalign {\vskip 0.4 em }
\noalign{\hrule depth0.005em}
\noalign{\vskip 0.08 em }
\noalign{\hrule depth0.005em}
\noalign {\vskip 4pt }
\omit \quad {\srfont All Energies are in MeV and Radii in fm}
\hidewidth \cr
\cr }} $$

\vfill\break
$$ \vbox { \offinterlineskip  \vskip6pt
\def\qq{\hskip 0.3em}

\def\titlestrut{\vrule depth4pt height10pt width0pt}
\def\upstrut{\vrule height10pt width0pt}
\def\downstrut{\vrule height12pt width0pt}
\halign {\qq#\hfil &&\qq# &\qq# &\qq\hfil#\hfil \cr
\multispan{12} \hfil {\bf TABLE 4. }{\bf Energy expectation values (calculated and fitted)} \hfil \titlestrut &\cr
\multispan{7} \hfil {\bf for $_{\Lambda}^4H$(Ground State) with {\bf $\bar V$=6.20} MeV} \hfil \titlestrut &\cr
\noalign {\vskip 0.5 em }
\noalign{\hrule depth0.005em}
\noalign{\vskip 0.08 em }
\noalign{\hrule depth0.005em}
{\hskip 2.5 em} $V_{\sigma}$ && $W_{o}$ && $C_{p}$ && $<E_{cal}>$ && $<E_{fit}>$ \downstrut &\cr
\noalign {\vskip 0.5 em }
\noalign{\hrule depth0.005em}
\noalign{\vskip 0.08 em }
\noalign{\hrule depth0.005em}
\noalign {\vskip 0.2 em }
{\hskip 2.15 em} 0.17 && 0 && 0 &&  10.58(05) && 10.60  \upstrut &\cr
\noalign {\vskip 0.15 em }
{\hskip 2.15 em} 0.17 && 0 && 1.0 && 11.61(07) && 11.63 \upstrut &\cr
\noalign {\vskip 0.15 em }
{\hskip 2.15 em} 0.17 && 0.01 && 0 &&  9.97(05) && 9.97 \upstrut &\cr
\noalign {\vskip 0.15em }
{\hskip 2.15 em} 0.17 && 0.01 && 1.0 &&  10.96(05) && 10.90 \upstrut &\cr
\noalign {\vskip 0.15em }
{\hskip 2.15 em} 0.17 && 0.02 && 0 &&  9.25(04) && 9.31 \upstrut &\cr
\noalign {\vskip 0.15em }
{\hskip 2.15 em} 0.17 && 0.02 && 1.0 && 10.21(05) && 10.13 \upstrut &\cr
\noalign {\vskip 0.15em }
{\hskip 2.15 em} 0.17 && 0.02 && 2.0 &&  12.28(07) && 12.31 \upstrut &\cr
\noalign {\vskip 0.15em }
{\hskip 2.15 em}  0.17 && 0.005 && 0 && 10.23(05) && 10.29 \upstrut &\cr
\noalign {\vskip 0.15em }
{\hskip 2.15 em} 0.17 && 0.005 && 1.0 && 11.25(05) && 11.27 \upstrut &\cr
\noalign {\vskip 0.15em }
{\hskip 2.15 em} 0.17 && 0.005 && 2.0 &&  13.58(08) && 13.60 \upstrut &\cr
\noalign {\vskip 0.15em }
{\hskip 2.15 em}  0.17 && 0.015 && 1.0 &&  10.50(05) && 10.52 \upstrut &\cr
\noalign {\vskip 0.15em }
{\hskip 2.15 em} 0.22 && 0 && 0 && 10.82(05) && 10.71 \upstrut &\cr
\noalign {\vskip 0.15em }
{\hskip 2.15 em}  0.22 && 0 && 1.0 &&  11.66(05) && 11.74 \upstrut &\cr
\noalign {\vskip 0.15em }
{\hskip 2.15 em} 0.22 && 0.01 &&  0 && 10.02(05) && 10.09 \upstrut &\cr
\noalign {\vskip 0.15em }
{\hskip 2.15 em} 0.22 && 0.01 && 1.0 &&  10.90(05) && 11.01 \upstrut &\cr
\noalign {\vskip 0.15em }
{\hskip 2.15 em} 0.22 && 0.01 && 2.0 && 13.36(08) && 13.30 \upstrut &\cr
\noalign {\vskip 0.15em }
{\hskip 2.15 em} 0.22 && 0.02 && 0 && 9.43(05) &&  9.43 \upstrut &\cr
\noalign {\vskip 0.15em }
{\hskip 2.15 em} 0.22 && 0.02 && 1.0 &&  10.21(05) && 10.25 \upstrut &\cr
\noalign {\vskip 0.15em }
{\hskip 2.15 em} 0.22 && 0.02 && 2.0 &&  12.35(06) && 12.43 \upstrut &\cr
\noalign {\vskip 0.15em }
{\hskip 2.15 em} 0.22 && 0.005 && 1.0 &&  11.47(05) && 11.38 \upstrut &\cr
\noalign {\vskip 0.15em }
{\hskip 2.15 em} 0.22 && 0.015 && 1.0 &&  10.68(05) && 10.63 \upstrut &\cr
\noalign {\vskip 0.15em }
{\hskip 2.15 em} 0.27 && 0 && 0 && 10.79(05) && 10.83 \upstrut &\cr
\noalign {\vskip 0.15em }
{\hskip 2.15 em} 0.27 && 0 && 1.0 &&  11.79(06) && 11.86 \upstrut &\cr
\noalign {\vskip 0.15em }
{\hskip 2.15 em} 0.27 && 0 && 2.0 &&  14.32(08) && 14.24 \upstrut &\cr
\noalign {\vskip 0.15em }
{\hskip 2.15 em} 0.27 && 0.01 && 0 &&  10.26(05) && 10.20 \upstrut &\cr
\noalign {\vskip 0.15em }
{\hskip 2.15 em} 0.27 && 0.01 && 1.0 && 11.13(05) && 11.12 \upstrut &\cr
\noalign {\vskip 0.15em }
{\hskip 2.15 em} 0.27 && 0.01 && 2.0 && 13.41(08) && 13.41 \upstrut &\cr
\noalign {\vskip 0.15em }
{\hskip 2.15 em} 0.27 && 0.02 && 0 && 9.60(05) && 9.55 \upstrut &\cr
\noalign {\vskip 0.15em }
{\hskip 2.15 em} 0.27 && 0.02 && 1.0 && 10.39(05) && 10.36 \upstrut &\cr
\noalign {\vskip 0.15em }
{\hskip 2.15 em} 0.27 && 0.02 && 2.0 && 12.56(06) && 12.54 \upstrut &\cr
\noalign {\vskip 0.15em }
{\hskip 2.15 em} 0.27 && 0.005 && 0 && 10.50(05) && 10.52 \upstrut &\cr
\noalign {\vskip 0.15em }
{\hskip 2.15 em} 0.27 && 0.005 && 1.0 && 11.46(05) && 11.49 \upstrut &\cr
\noalign {\vskip 0.15em }
{\hskip 2.15 em} 0.27 &&  0.015 && 1.0 && 10.81(05) && 10.75 \upstrut &\cr
\noalign {\vskip 0.4em }
\noalign{\hrule depth0.005em}
\noalign{\vskip 0.08 em }
\noalign{\hrule depth0.005em}
\noalign {\vskip 4pt}
\omit \quad  $\epsilon=$0.24,  $\kappa_{\Lambda N}=$0.12,  $\alpha_s=$0.95,  $\alpha_{\sigma}=$0.7  \hidewidth  \cr
}} $$

$$ \vbox { \offinterlineskip  \vskip6pt
\def\qq{\hskip 0.3em}

\def\titlestrut{\vrule depth4pt height10pt width0pt}
\def\upstrut{\vrule height10pt width0pt}
\def\downstrut{\vrule height12pt width0pt}
\halign {\qq#\hfil &&\qq# &\qq\hfil#\hfil \cr
\multispan{16} \hfil {\bf TABLE 5. $B_{\Lambda}$ as a function of coefficients $y_{1-6}^{\dagger}$} \hfil \titlestrut &\cr
\noalign {\vskip 0.4 em }
\noalign {\hrule depth0.005em}
\noalign {\vskip 0.08 em }
\noalign {\hrule depth0.005em}
\hskip2.0 em ${_{\Lambda}^A}Z$ && ${\bar V}$ && $y_1$ && $y_2$ && $y_3$ && $y_4$ && $y_5$ && $y_6$ \upstrut &\cr
\noalign {\vskip 0.4 em }
\noalign {\hrule depth0.005em}
\noalign {\vskip 0.08 em }
\noalign {\hrule depth0.005em}
\hskip2.0em ${_{\Lambda}^5}He$ && 6.20&& -0.92 && -285.51 && 1.28 && 1701.78 && 1.50 && -37.62 \upstrut &\cr
\hskip0.2em && 6.15&& \hskip2pt 1.19&& -280.03 && 0.99 && 1134.41 && 1.35 && \hskip6pt -7.44 \upstrut &\cr
\hskip0.2em && 6.10&& \hskip2pt -0.20&& -210.02 && 0.88 && 2997.79 && 1.19 && \hskip6pt -44.03 \upstrut &\cr
\noalign {\vskip 0.6 em }
\hskip2.0em ${_{\Lambda}^4}H$ && 6.20&& \hskip2pt 2.27&& \hskip4pt -61.23&& 0.35&& \hskip7pt -141.55&& 0.68&& -10.53 \upstrut &\cr
\hskip0.2em && 6.15&& \hskip2pt 2.21&& \hskip4pt -59.03&& 0.32&& \hskip7pt 453.47&& 0.52&& \hskip6pt -8.14 \upstrut &\cr
\hskip0.2em && 6.10&& \hskip2pt 1.55&& \hskip4pt -43.48&& 0.23&& \hskip7pt 298.23&& 0.37&& \hskip6pt -4.96 \upstrut &\cr
\noalign {\vskip 0.6 em }
\hskip2.0 em ${_{\Lambda}^4}H$$^{*}$ && 6.20&& -0.89&& -112.62&& 0.37&& \hskip7pt 639.67&& 0.68&& \hskip6pt -8.51 \upstrut &\cr
\hskip0.2em && 6.15&& -0.98&& \hskip4pt -65.63&& 0.27&& \hskip7pt 208.58&& 0.45&& \hskip6pt -6.68 \upstrut &\cr
\hskip0.2em && 6.10&& -0.81&& \hskip4pt -53.21&& 0.12&& \hskip7pt 334.70&& 0.36&& \hskip6pt -5.69 \upstrut &\cr
\noalign {\vskip 0.6 em }
\noalign {\hrule depth0.005em}
\noalign {\vskip 0.08 em }
\noalign {\hrule depth0.005em}
\noalign {\vskip 2pt}
\omit \quad $\dagger$ {\srfont Includes contribution due to $\Lambda$-nuclear correlations} \hidewidth \cr
}} $$

$$ \vbox { \offinterlineskip  \vskip6pt
\def\qq{\hskip 0.3em}

\def\titlestrut{\vrule depth4pt height10pt width0pt}
\def\upstrut{\vrule height10pt width0pt}
\def\downstrut{\vrule height12pt width0pt}
\halign {\qq#\hfil &&\qq# &\qq\hfil#\hfil \cr
\multispan{7} \hfil {\bf TABLE 6(A).}$\,$ {\bf Variational Results for {\bf ${_{\Lambda}^5}He$}  } \hfil \titlestrut &\cr
\noalign {\vskip 0.25 em }
\noalign{\hrule depth0.005em}
\noalign{\vskip 0.08 em }
\noalign{\hrule depth0.005em}
\hskip2.0 em Components && \hskip 0.2 truecm $\bar V= 6.20 $ && \hskip 1.0 truecm $\bar V= 6.15 $ && \hskip 0.5 truecm $\bar V= 6.10 $ \upstrut &\cr
\noalign {\vskip 0.4 em }
\noalign{\hrule depth0.005em}
\noalign{\vskip 0.08 em }
\noalign{\hrule depth0.005em}
\noalign {\vskip 0.4 em }
\hskip0.2 em Nuclear Kinetic energy$^{\dagger}$ && \hskip0.3 truecm 128.38(74) && \hskip1.1 truecm 128.16(71) && \hskip0.55 truecm 125.35(69) \upstrut &\cr
\hskip0.2 em {\it NN} Potential energy$^{\dagger}$ && \hskip0.2 truecm -139.97(73) && \hskip1.0 truecm -140.27(69) && \hskip0.4 truecm -139.31(67) \upstrut &\cr
\hskip0.2 em {\it NNN} Potential energy && \hskip0.8 truecm -5.97(8) && \hskip1.6 truecm -5.91(8) && \hskip1.05 truecm -5.73(8) \upstrut &\cr
\hskip0.2 em $\Lambda$ Kinetic energy && \hskip0.6 truecm 11.64(15) && \hskip1.3 truecm 11.11(15) && \hskip1.05 truecm 9.13(13) \upstrut &\cr
\hskip0.2 em $\Lambda$$N$ P.E (central) && \hskip0.5 truecm -23.65(32) && \hskip1.2 truecm -21.69(30) && \hskip0.7 truecm -17.55(27) \upstrut &\cr
\hskip0.2 em $\Lambda${\it N} P.E (spin) && \hskip0.5 truecm -0.0138(1) && \hskip1.2 truecm -0.0311(3) && \hskip0.7 truecm -0.0281(3) \upstrut &\cr
\hskip0.2 em $\Lambda${\it N} Space exch. contribution && \hskip0.6 truecm 0.763(13) && \hskip1.4 truecm 0.578(10) && \hskip1.05 truecm 0.395(7) \upstrut &\cr
\hskip0.2 em $\Lambda${\it NN} P.E (Total) && \hskip0.9 truecm -1.91(9) && \hskip1.4 truecm -2.67(10) && \hskip0.9 truecm -3.53(10) \upstrut &\cr
\hskip0.2 em $\Lambda${\it NN} P.E (T.P.E.) && \hskip0.7 truecm -8.43(16) && \hskip1.4 truecm -8.77(17) && \hskip0.85 truecm -8.20(17) \upstrut &\cr
\hskip0.2 em $\Lambda${\it NN} P.E (dispersive) && \hskip0.8 truecm 6.52(11) && \hskip1.5 truecm 6.10(11) && \hskip1.0 truecm 4.67(10) \upstrut &\cr
\hskip0.2 em {\bf Total Energy} && \hskip0.2 truecm {\bf -30.75(14)} && \hskip0.9 truecm {\bf -30.76(18)} && \hskip0.4 truecm {\bf -31.27(12)} \upstrut &\cr
\hskip0.2 em {\bf B$_{\Lambda}$} && \hskip0.6 truecm {\bf 3.03(15)} && \hskip1.3 truecm {\bf 3.05(19)} && \hskip0.75 truecm {\bf 3.56(13)} \upstrut &\cr
\noalign {\vskip 0.4 em }
\hskip0.2 em r.m.s. radius (proton) && \hskip0.8 truecm 1.376(2) && \hskip1.55 truecm 1.379(2) && \hskip0.95 truecm 1.389(2) \upstrut &\cr
\hskip0.2 em r.m.s. radius (neutron) && \hskip0.75 truecm 1.377(2) && \hskip1.5 truecm 1.379(2) && \hskip0.95 truecm 1.389(2) \upstrut &\cr
\hskip0.2 em d-state probability && \hskip0.55 truecm 0.1568(2) && \hskip1.3 truecm 0.1568(2) && \hskip0.7 truecm 0.1557(2) \upstrut &\cr
\noalign {\vskip 0.6 em }
\noalign{\hrule depth0.005em}
\noalign{\vskip 0.08 em }
\noalign{\hrule depth0.005em}
\noalign{\vskip2pt}
\omit \quad $\dagger$ {\srfont Includes contribution due to $\Lambda$-nuclear correlations} \hidewidth \cr
}} $$

\vfill\break

$$ \vbox { \offinterlineskip  \vskip6pt
\def\qq{\hskip 0.3em}

\def\titlestrut{\vrule depth4pt height10pt width0pt}
\def\upstrut{\vrule height10pt width0pt}
\def\downstrut{\vrule height12pt width0pt}
\halign {\qq#\hfil &&\qq# &\qq\hfil#\hfil \cr
\multispan{7} \hfil {\bf TABLE 6(B).}$\,$ {\bf Variational results for {\bf ${_{\Lambda}^4}H$}} \hfil \titlestrut &\cr
\noalign {\vskip 0.25 em }
\noalign{\hrule depth0.005em}
\noalign{\vskip 0.08 em }
\noalign{\hrule depth0.005em}
\hskip2.0 em Components && \hskip 0.2 truecm $\bar V= 6.20$ && \hskip 1.0 truecm $\bar V= 6.15$ && \hskip 1.0 truecm $\bar V= 6.10$ \upstrut &\cr
\noalign {\vskip 0.4 em }
\noalign{\hrule depth0.005em}
\noalign{\vskip 0.08 em }
\noalign{\hrule depth0.005em}
\noalign {\vskip 0.4 em }
Nuclear Kinetic energy$^{\dagger}$ && \hskip0.3 truecm 65.76(47) && \hskip1.15 truecm 62.82(47) && \hskip1.15 truecm 60.34(46) \upstrut &\cr
{\it NN} Potential energy$^{\dagger}$ && \hskip0.2 truecm -66.91(47) && \hskip1.0 truecm -65.10(46) && \hskip1.0 truecm -63.60(46) \upstrut &\cr
{\it NNN} Potential energy && \hskip0.6 truecm -1.33(3) && \hskip1.4 truecm -1.25(3) && \hskip1.45 truecm -1.19(3) \upstrut &\cr
$\Lambda$ Kinetic energy && \hskip0.55 truecm 7.17(11) && \hskip1.6 truecm 5.99(9) && \hskip1.55 truecm 4.77(9) \upstrut &\cr
$\Lambda${\it N} P.E (central) && \hskip0.2 truecm -14.00(22) && \hskip1.05 truecm -11.55(20) && \hskip1.25 truecm -8.84(18) \upstrut &\cr
$\Lambda${\it N} P.E (spin) && \hskip0.4 truecm -0.291(3) && \hskip1.25 truecm -0.363(4) && \hskip1.3 truecm -0.300(4) \upstrut &\cr
$\Lambda${\it N} Space exch. contribution && \hskip0.5 truecm 0.333(9) && \hskip1.4 truecm 0.221(6) && \hskip1.45 truecm 0.143(4) \upstrut &\cr
$\Lambda${\it NN} P.E (Total) && \hskip0.6 truecm -1.33(6) && \hskip1.5 truecm -1.36(6) && \hskip1.5 truecm -1.48(6) \upstrut &\cr
$\Lambda${\it NN} P.E (T.P.E.) && \hskip0.6 truecm -2.87(8) && \hskip1.5 truecm -2.63(9) && \hskip1.5 truecm -2.47(9) \upstrut &\cr
$\Lambda${\it NN} P.E (dispersive) && \hskip0.75 truecm 1.54(4) && \hskip1.6 truecm 1.27(4) && \hskip1.6 truecm 0.99(3) \upstrut &\cr
{\bf Total Energy} && \hskip0.2 truecm {\bf -10.61(6)} && \hskip1.0 truecm {\bf -10.59(5)} && \hskip1.0 truecm {\bf -10.16(5)} \upstrut &\cr
{\bf B$_{\Lambda}$} && \hskip0.6 truecm {\bf 2.28(6)} && \hskip1.4 truecm {\bf 2.27(6)} && \hskip1.4 truecm {\bf 1.83(6)} \upstrut &\cr
\noalign {\vskip 0.4 em }
r.m.s. radius (proton) && \hskip0.55 truecm 1.408(2) && \hskip1.4 truecm 1.435(3) && \hskip1.4 truecm 1.460(3) \upstrut &\cr
r.m.s. radius (neutron) && \hskip0.55 truecm 1.516(3) && \hskip1.4 truecm 1.547(3) && \hskip1.4 truecm 1.577(3) \upstrut &\cr
d-state probability && \hskip0.3 truecm 0.0984(1) && \hskip1.2 truecm 0.0960(1) && \hskip1.2 truecm 0.0962(1) \upstrut &\cr
\noalign {\vskip 0.6 em }
\noalign{\hrule depth0.005em}
\noalign{\vskip 0.08 em }
\noalign{\hrule depth0.005em}
\noalign{\vskip2pt}
\omit \quad $\dagger$ {\srfont Includes contribution due to $\Lambda$-nuclear correlations} \hidewidth \cr
}} $$

$$ \vbox { \offinterlineskip  \vskip6pt
\def\qq{\hskip 0.3em}

\def\titlestrut{\vrule depth4pt height10pt width0pt}
\def\upstrut{\vrule height10pt width0pt}
\def\downstrut{\vrule height12pt width0pt}
\halign {\qq#\hfil &&\qq# &\qq\hfil#\hfil \cr
\multispan{7} \hfil {\bf TABLE 6(C).}$\,$ {\bf Variational results for {\bf ${_{\Lambda}^4}H$$^{*}$} } \hfil \titlestrut &\cr
\noalign {\vskip 0.25 em }
\noalign{\hrule depth0.005em}
\noalign{\vskip 0.08 em }
\noalign{\hrule depth0.005em}
\hskip2.0 em Components && \hskip 0.2 truecm $\bar V= 6.20 $ && \hskip 1.0 truecm $\bar V= 6.15 $ && \hskip 1.0 truecm $\bar V= 6.10 $ \upstrut &\cr
\noalign {\vskip 0.4 em }
\noalign{\hrule depth0.005em}
\noalign{\vskip 0.08 em }
\noalign{\hrule depth0.005em}
\noalign {\vskip 0.4 em }
Nuclear Kinetic energy$^{\dagger}$ && \hskip0.35 truecm 63.78(49) && \hskip1.15 truecm 60.58(47) && \hskip1.15 truecm 57.96(46) \upstrut &\cr
{\it NN} Potential energy$^{\dagger}$ && \hskip0.2 truecm -65.71(48) && \hskip1.0 truecm -63.64(46) && \hskip1.0 truecm -62.36(45) \upstrut &\cr
{\it NNN} Potential energy && \hskip0.6 truecm -1.34(3) && \hskip1.4 truecm -1.27(3) && \hskip1.45 truecm -1.20(3) \upstrut &\cr
$\Lambda$ Kinetic energy && \hskip0.5 truecm 6.40(12) && \hskip1.5 truecm 4.78(9) && \hskip1.2 truecm 3.4516(8) \upstrut &\cr
$\Lambda${\it N} P.E (central) && \hskip0.15 truecm -12.35(25) && \hskip1.2 truecm -9.42(19) && \hskip1.25 truecm -6.56(17) \upstrut &\cr
$\Lambda${\it NN} P.E (spin) && \hskip0.5 truecm 0.086(1) && \hskip1.3 truecm 0.083(1) && \hskip1.4 truecm 0.065(1) \upstrut &\cr
$\Lambda${\it N} Space exch. contribution && \hskip0.5 truecm 0.311(8) && \hskip1.3 truecm 0.202(6) && \hskip1.4 truecm 0.110(4) \upstrut &\cr
$\Lambda${\it NN} P.E (Total) && \hskip0.5 truecm -0.57(6) && \hskip1.3 truecm -0.62(5) && \hskip1.5 truecm -0.73(5) \upstrut &\cr
$\Lambda${\it NN} P.E (T.P.E.) && \hskip0.5 truecm -3.11(9) && \hskip1.3 truecm -2.32(8) && \hskip1.5 truecm -1.98(8) \upstrut &\cr
$\Lambda${\it NN} P.E (dispersive) && \hskip0.65 truecm 2.54(7) && \hskip1.45 truecm 1.70(5) && \hskip1.65 truecm 1.24(5) \upstrut &\cr
{\bf Total Energy} && \hskip0.3 truecm {\bf -9.40(7)} && \hskip1.1 truecm -{\bf 9.31(5)} && \hskip1.3 truecm {\bf -9.26(5)} \upstrut &\cr
{\bf B$_{\Lambda}$} && \hskip0.4 truecm {\bf 1.08(7)} && \hskip1.25 truecm {\bf 0.99(5)} && \hskip1.45 truecm {\bf 0.94(5)} \upstrut &\cr
\noalign {\vskip 0.4 em }
r.m.s. radius (proton) && \hskip0.4 truecm 1.431(3) && \hskip1.3 truecm 1.467(3) && \hskip1.45 truecm 1.494(3) \upstrut &\cr
r.m.s. radius (neutron) && \hskip0.4 truecm 1.542(3) && \hskip1.3 truecm 1.585(3) && \hskip1.45 truecm 1.618(3) \upstrut &\cr
d-state probability && \hskip0.2 truecm 0.0995(1) && \hskip1.1 truecm 0.0980(1) && \hskip1.2 truecm 0.0969(1) \upstrut &\cr
\noalign {\vskip 0.6 em }
\noalign{\hrule depth0.005em}
\noalign{\vskip 0.08 em }
\noalign{\hrule depth0.005em}
\noalign{\vskip2pt}
\omit \quad $\dagger$ {\srfont Includes contribution due to $\Lambda$-nuclear correlations} \hidewidth \cr
}} $$

$$ \vbox { \offinterlineskip  \vskip6pt
\def\qq{\hskip 0.3em}

\def\titlestrut{\vrule depth4pt height10pt width0pt}
\def\upstrut{\vrule height10pt width0pt}
\def\downstrut{\vrule height12pt width0pt}
\halign {\qq#\hfil &&\qq# &\qq\hfil#\hfil \cr
\multispan{15} \hfil {\bf TABLE 7.} {\bf Breakup of the ${0^{+}-1^{+}}$ splitting contributions} \hfil \titlestrut &\cr
\noalign {\vskip 0.5 em }
\noalign{\hrule depth0.005em}
\noalign{\vskip 0.08 em }
\noalign{\hrule depth0.005em}
\hskip 0.8 em Contribution && \hskip 0.8 em $\bar V=6.20$ && \hskip 0.5 truecm $\bar V=6.15$ && \hskip 0.7 truecm $\bar V=6.10$  \upstrut &\cr
\noalign {\vskip 0.4 em }
\noalign{\hrule depth0.005em}
\noalign{\vskip 0.08 em }
\noalign{\hrule depth0.005em}
\noalign {\vskip 0.4 em }
\hskip 1.0 em $V_{\sigma}$ && \hskip0.55 truecm 0.377(3) && \hskip0.7 truecm 0.446(4) && \hskip0.9 truecm 0.365(4) \upstrut &\cr
\hskip 1.0 em $V_{\Lambda NN}$ && \hskip0.8 truecm 0.76(8) && \hskip0.9 truecm 0.74(8) && \hskip1.1 truecm 0.75(8) \upstrut &\cr
\hskip 1.0 em Total && \hskip0.4 truecm 1.137(80) && \hskip0.45 truecm 1.186(80) && \hskip0.7 truecm  1.115(89) \upstrut &\cr
\hskip 1.0 em Energy differences && \hskip0.8 truecm 1.21(9) && \hskip0.9 truecm 1.28(7) && \hskip1.1 truecm  0.90(7) \upstrut &\cr
\noalign {\vskip 0.6 em }
\noalign{\hrule depth0.005em}
\noalign{\vskip 0.08 em }
\noalign{\hrule depth0.005em}
\noalign{\vskip4pt}
\omit \quad {\srfont The first row gives the contribution to splitting from V$_{\sigma}$. The second row gives the } \hidewidth \cr
\omit \quad {\srfont contribution arising from V$_{\Lambda NN}$. The third row gives the total of V$_{\sigma}$ and V$_{\Lambda NN}$.} \hidewidth \cr
\omit \quad {\srfont The last row gives the actual calculated energy difference between ${_{\Lambda}^4}H$ and ${_{\Lambda}^4}H$$^{*}$.} \hidewidth
\cr }}
$$

$$ \vbox { \offinterlineskip  \vskip6pt
\def\qq{\hskip 0.3em}

\def\titlestrut{\vrule depth4pt height10pt width0pt}
\def\upstrut{\vrule height10pt width0pt}
\def\downstrut{\vrule height12pt width0pt}
\halign {\qq#\hfil &&\qq# &\qq\hfil#\hfil \cr
\multispan{12} \hfil {\bf TABLE 8.}$\, \,${\bf Results for Nuclear Matter calculations} \hfil \titlestrut &\cr
\noalign {\vskip 0.5 em }
\noalign{\hrule depth0.005em}
\noalign{\vskip 0.08 em }
\noalign{\hrule depth0.005em}
{\hskip 1.4 em} $\bar V$ && $\rho_{o}$ && -$D$ && $<T_\Lambda+V_{\Lambda N}>$ && Spc. exch. && $V_{\Lambda NN}$ \downstrut &\cr
\noalign {\vskip 0.4 em }
\noalign{\hrule depth0.005em}
\noalign{\vskip 0.08 em }
\noalign{\hrule depth0.005em}
\noalign {\vskip 0.4 em }
{\hskip 0.75 em} 6.20 && 0.162 && -21.525 && -77.617 && 8.561 && 47.530 \upstrut &\cr
\noalign {\vskip 0.5 em }
{\hskip 0.75 em} 6.15 && 0.162 && -17.819 && -72.628 && 6.644 && 48.164 \upstrut &\cr
\noalign {\vskip 0.5 em }
{\hskip 0.75 em} 6.10 && 0.162 && -11.727 && -67.729 && 4.798 && 51.205 \upstrut &\cr
\noalign {\vskip 0.6 em }
\noalign{\hrule depth0.005em}
\noalign{\vskip 0.08 em }
\noalign{\hrule depth0.005em}
\noalign {\vskip 0.4 em }
\noalign{\vskip4pt}
\omit \quad {\srfont All energies are in MeV. $\rho_{o}$ is the normal nuclear matter density in fm$^{-3}$}. \hidewidth \cr
\omit \quad {\srfont The third column gives the well depth {\it D}. The fourth column gives the value } \hidewidth \cr
\omit \quad {\srfont  of {\it D} without the $\Lambda${\it NN} forces and the space-exchange contribution, i.e. for } \hidewidth \cr
\omit \quad {\srfont $\epsilon=0$. The fifth column gives the reduction in the contribution to {\it D} due to} \hidewidth \cr
\omit \quad {\srfont the space-exchange part (Spc. exch.) of the $\Lambda${\it N} potential. The last column  } \hidewidth \cr
\omit \quad {\srfont gives the contribution due to the three-body $\Lambda${\it NN} forces.} \hidewidth
\cr }}
$$

\end